# Bridging the Physics and Chemistry of Graphene(s): From Hückel's Aromaticity to Dirac's Cones and Topological Insulators

Aristides D. Zdetsis*

Molecular Engineering Laboratory, Department of Physics, University of Patras, Patras 26500 GR, Greece

## ABST RACT

By bridging graphene and benzene through a well-defined sequence of polycyclic aromatic hydrocarbons and their inherent shell structure, it is shown that graphene is actually a coherent arrangement of interwoven benzene molecules, coordinated by aromaticity, shell structure, and topology, all interrelated and microscopically realized through dynamical flipping of the atomic $p_z$-orbitals, playing the role of pseudospin or "qubits". This renders graphene resonance structure, "resonating" between two complementary aromaticity patterns, involving $2^k$, k→∞ kekulé type of resonances resulting in "robust electronic coherence", with dual "molecular-crystalline" nature, and two valence-conduction bands of opposite parity, driven by inversion symmetry competition, which is essentially a "molecule-versus-crystal" competition, in accord with topological-insulator and many-body theory. The "average picture" converges to the usual band structure with two aromatic π-electrons per ring, and the fingerprints of inversion-competition at the $D_{3h}$-symmetric Dirac points, which for rectangular nanographene(s) appear as gapless topological edge states without real spin-polarization, contrary to opposite claims.

Email:zdetsis@upatras.gr

**TOC Graphic**

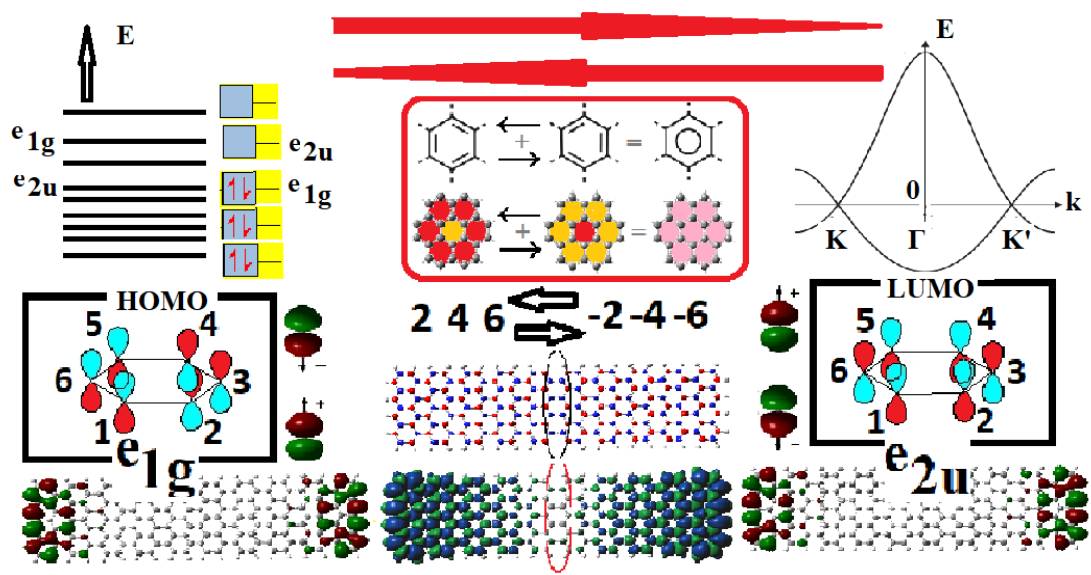

**1.    Introduction** The molecular or Chemical description of graphene (in real space) has been largely overlooked in the literature compared to the "crystalline" or Physical picture (in k-space), which are clearly interrelated (but not in a simple or direct way). Bridging the two descriptions (bonding and "banding") is best accomplished by bridging benzene and graphene through a sequence of polycyclic aromatic hydrocarbons (PAHs) of growing size all the way to infinity. Strangely enough, the same process of bridging the molecular and crystalline aspects of graphene is shown here to be the underlying natural process responsible for all exotic properties of graphene, which at the deepest level are rooted in the competition between the molecular and crystalline nature of graphene. The vehicle for such process is aromaticity (of benzene) coupled with the topological (symmetry) properties of the honeycomb lattice, both incorporated (directly or indirectly) in the "shell model".[1]   This process is illustrated through a sequence of hexagonal PAHs, called for brevity the "main sequence"[1] which is generated by the shell model[1] as the number of shells n increases, presumably all the way to infinity.  It is shown, in an transparent, straightforward, and insightful way, that all unusual and exotic characteristics of graphene and the uncommon and abstract concepts used in their description[1-7] (such as Dirac points and Dirac  cones, Berry phases, topological insulators etc) are related to topological and symmetry constrains which are ultimately generated by the competition of molecular symmetry ($D_{6h}$) against crystalline (or sublattice) $D_{3h}$ symmetry, with and without inversion symmetry,  respectively.  This includes features going well beyond the common band structure, such as the "robust electronic coherence"[4] of graphene. It is well known that although the band structure of graphene can successfully describe most of the experimental data, the subject of the electron-electron Coulomb interactions is still active and open to further investigation.[4-7] Although the present  approach is based on the "mean field" one-body density functional theory (DFT) approximation, is in fact indirectly incorporating many-body effects through many (presumably infinite) "one-body" interconnected, and interrelated calculations on sequentially larger size PAHs of the same hexagonal symmetry. Such calculations performed in a systematic well-defined way are fully, smoothly, and well converging to the electronic characteristics of graphene, both "static" (band properties) and dynamic (many-body/topological fluctuations).This is because, as was stated earlier, the present real space ("molecular") approach incorporates in a simple and efficient way (and emphasizes) the key feature of crystalline-versus-molecular symmetry competition, which turns out to be the driving force towards the electronic coherence and similar effects, described by many body theory and other advanced techniques[5-8]. Thus, this molecular approach is proven capable of reproducing (naturally, and practically effortlessly) at least qualitatively but accurately, key advanced properties, such as "robust electronic coherence" of Dirac electrons, connected with electron-electron Coulomb interactions, usually obtained by many-body and other high-level theoretical techniques[5-8], together with "static" band-structure results e.g. Dirac's cones[4] at the $D_{3h}$ symmetric $\mathbf{K}$ and $\mathbf{K'}$ points at the edges of the Brillouin zones, and the electron-hole symmetry, both related to the shell-structure and the bipartite nature of the honeycomb "lattice", which is built gradually and indirectly as the size of the samples increases (with different sublattice sites corresponding to sites of different chirality, and different parity of the frontiers orbitals), see Fig. 1 below. In addition, besides the full understanding, interpretation, and unification of known (through band theory, many-body theory, and other advanced methods) data, totally new results and novel insight have been obtained especially on the key role of the atomic $p_z$ orbitals, and the topological edge states, together with aromaticity and topology (symmetry), all interrelated. The orientation of $p_z$ orbitals at each carbon site plays the role of pseudospin or qubits in an abstract resonating generalized-valence- bond state, not as yet mathematically formulated. The topological edge states on the other hand, which are totally missing in band theory due to the periodic boundary conditions, are shown to be largely misunderstood or misinterpreted  as spin polarized in most real space finite size atomistic (non-band) approaches, in particular for rectangular nanograhenes (NGRs) and graphene nanoribbons (GNRs) of both  armchair (AGNRs) and zigzag (ZGNRs) sides. Failure to recognize the true nature of such edge states could lead to additional misconceptions and discrepancies about the (finite) gap of AGNRs (and in part ZGNRs), *vide infra*.  The resulting "grand picture" not only unifies the "physics and chemistry of graphene" (in real and k-space), or the molecular versus crystalline characteristics of graphene  as the title states, but also fully interrelates the Physics and Chemistry of benzene with the Physics and Chemistry of Graphene (and the intermediate PAHs as well); revealing at the same time their strong similarity. At the basis of this work is the "shell-model",[1] which in fact contains and expands the fundamental, but semiempirical Hückel's and Clar's rules of aromaticity, and aromaticity itself (not as an input, but rather as a result). Aromaticity, which basically means "like benzene",[2] is in fact the emerging key "molecular" concept, which provides new and novel information and insight not only for the properties of infinite graphene, but also for the intermediate PAHS. Indeed, it can be clearly verified (see for instance the frontier orbitals of  successive PAHs in Fig.1 below) that graphene itself (and in part the intermediate PAHs) is not simply "like benzene", but essentially is and  behaves as a coherent and well-coordinated ensemble of interlinked and interlocked  benzene molecules behaving very much like a gigantic "super benzene", whose frontier orbitals (valence and conduction bands) are successive linear combinations of the frontier orbitals of all benzene rings comprising graphene (see Fig. S1, and relations S1, S2 and S1', S2').  The combined application of the shell model and geometrical/topological principles, related with sublattice (or crystal) versus molecular group symmetries, is the underlying reason(s) for most of the "exotic" properties of graphene, which can be essentially considered as macroscopic manifestations of aromaticity. Although, aromaticity of graphene,[2, 8-10] and AGNRs[10-13] has been previously considered in several respects, mostly at the phenomenological level (using small "isoelectronic" molecular models, or the celebrated, but empirical, Clar's rules), there was not up to now (according to the present author's knowledge) a fundamental, or unifying microscopic *ab initio* understanding (in depth and breadth) of the deeper relationship of aromaticity and the "exotic" properties of graphene. As is illustrated here, graphene is not simply aromatic, but is and behaves as a crystalline (macroscopic) prototype of aromaticity in the same way as benzene is a molecular prototype of aromaticity. A fundamental "by-product" of the present study, as mentioned earlier, is the deeper interrelation of aromaticity and topology which leads to topological edge states and topological validation of the "empirical" Clar's and Hückel's rules through shell closure(s).[1] It should be emphasized here that all "advanced" concepts used in the theoretical description of graphene, which have been introduced in the simplest possible way, are (by definition) described in k-space language. Therefore, their "molecular" analogues, or their molecular roots, should not be expected to be defined in the same mathematical way (nor in one-to-one correspondence), but should be conceptually "similar". This is the general plan of the present paper. The methodology followed here has evolved in recent years through successive (and successful) attempts made by the present author and collaborators[1-2, 11-12] to approach graphene sequentially or "dynamically" with the guidelines (on top of proven efficiency) of: 1) simplicity, 2) transparency, and 3) reproducibility; avoiding as much as possible complicated and/or overspecialized mathematical methods and tools, or controversial techniques and approaches; obviously under the provision that the desired results can be safely obtained by the simplest methods chosen (used in a resourceful, imaginative and productive way). Therefore, only simple symmetry arguments and notions are used from the point of view of topology (e.g. topological edge states); and similarly for aromaticity, only the most widespread and well known magnetic criterion of aromaticity was used based on the nucleus independent chemical shift, NICS(1)[14] (see methods), which was proven adequate and reliable[1-2, 12-15] for such study. Needless to say, that the concept of aromaticity, although fundamental and "multidimensional", is still considered controversial, not well-defined, and often misused or overused[16-17], with a large (ever growing) number of "aromaticity indices"[17] not always compatible with each other[2, 15, 17] and/or experiment. The present author, who is neither an aromaticity specialist nor a topology expert, believes that simplicity (and transparency) is an asset, not a drawback; therefore, both aspects are presented in the simplest, most fundamental, transparent, and easy to reproduce way. The hexagonal bridge to graphene through the "main sequence" of PAHs, underlining the shell model[1] is presented in section 2, where it is shown that the shell structure coupled with inversion symmetry breaking in the trigonal ($D_{3h}$) sublattice(s) of the hexagonal ($D_{6h}$) honeycomb lattice, is responsible for the puzzling behaviour of the main sequence PAHs and (hexagonal) graphene(s). It also explains why there are only two distinct aromaticity patterns (see Fig.1 below), topologically complementary to each other (as many as the sublattices) in hexagonal PAHs (the number of which for rectangular samples, as shown in section 3, becomes three). At the molecular scale the transformation from one pattern to the other is taking place through a flipping of the $p_z$ atomic orbitals in the two sublattices of graphene. For rectangular graphene(s), NGRs, and GNRs, the inversion symmetry breaking associated with crystalline (sublattice) versus molecular symmetry conflict leads to more novel results, such as the appearance of gapless topological edge states, which are largely missing in hexagonal PAHs and NGRs. This is described in section 3; while the key conclusions of the present study are compiled in section 4, and a brief description of the theoretical and computational framework is summarized in section 5.

## 2. The Hexagonal Bridge to Graphene: Results and Discussion for Hexagonal Samples.

The "main sequence" of PAHs shown in Fig.1, contains hexagonal PAHs with the general formula $C_{6n^2}H_{6n}$, $n = 1, 2, \cdots$ (n=1-7 in Fig.1) consisting of $n$ hexagonal monocyclic rings surrounding each other in a form of $n$ "babushka"-like Russian dolls. This "main sequence" defines the shell model or the "shell structure"[1] (summarized in section #S2 of supplementary information), which can be visualized by focusing in anyone of these PAHs with given n, and recognizing the inside layers as the preceding PAHs (n-1, n-2, …1), which constitute a full shell structure (geometric and electronic) very much analogous to the atomic shell structure underlying the periodical system of the elements.[1]

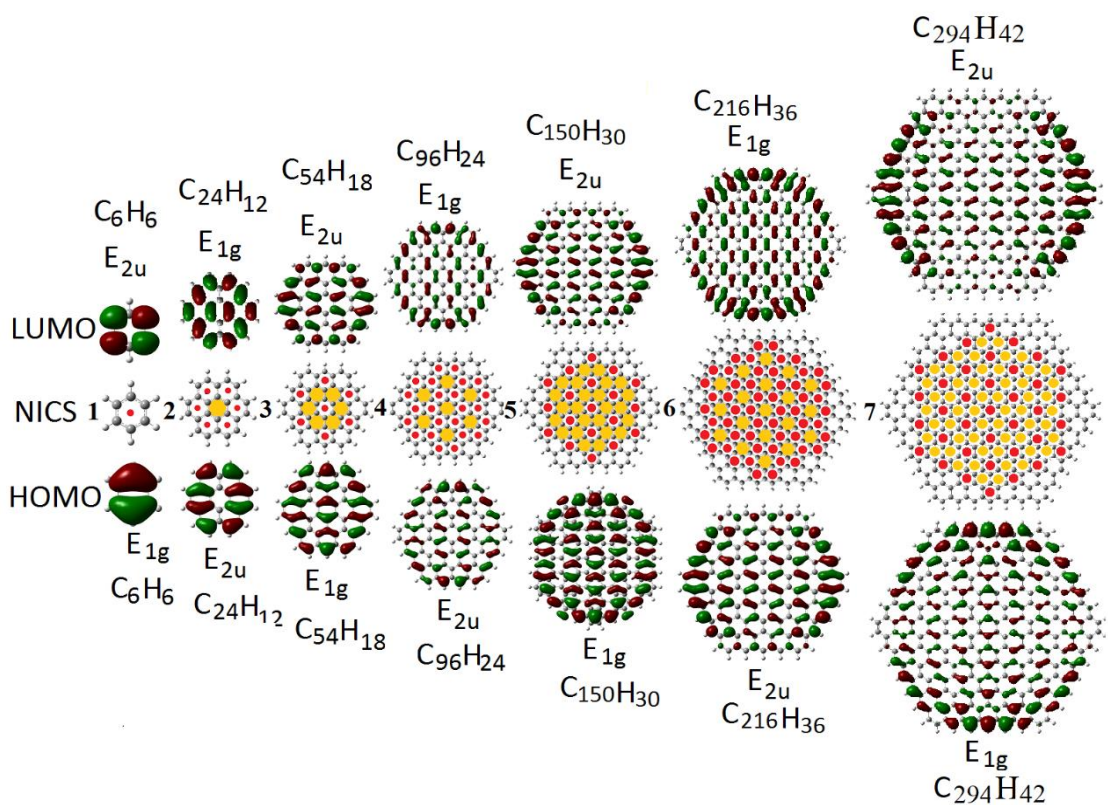

**FIGURE 1.** The first seven members and their stoichiometry of the "main sequence" PAHs bridging benzene to graphene, which are characterized by the "shell number" *n* (*n*=1-7). The aromaticity patterns based on the NICS(1) aromaticity index[10], describing the aromatic (or "full") rings are given with red (on line) dots at the centers of the rings; whereas the non-aromatic (or "empty") rings are shown for emphasis with yellow color (on line). The frontier MOs are shown on the lower (HOMO) and upper (LUMO) parts of the figure, together with their symmetry labels (representations).

There is one important lesson (among others) to be learned from Fig. 1, which is the periodicity of four interrelated fundamental characteristics. As the shell number n increases, alternating between odd and even values several (four) other physical quantities alternate as well:

**(a)** the symmetry of the highest occupied and of the lowest unoccupied (HOMO, and LUMO respectively) molecular orbitals (MOs) alternates also between even and odd parity (*e.g.*

see also Fig. S1). This can be clearly understood from the shell structure and the fact that each layer addition adds 2n+1 2D MOs carrying (2n+1)×4 ("2D") electrons ( and 2n+1 1D MOs carrying (2n+1)×2 "1D" electrons)[1]. The new HOMO and LUMO MOs will come from the new 2n+1 2D occupied orbitals, which are of two alternating types (with odd and even parity). Therefore, if the previous HOMO (and LUMO) is of one type, the new HOMO (and LUMO) would be of the opposite type.

**(b)** The aromatic ('full") and non-aromatic ("empty") rings, shown with red and yellow circles, respectively at their centers, alternate as well. This in fact constitutes a visualization and verification of the shell relations S1, S2 and S1', S2' and is a direct consequence of the previous HOMO-LUMO alternation (a).

**(c)** As will be illustrated below, the above interchanges are also accompanied by an interchange (and/or "coupling") of sublattice sites A and B, realized as an interchange (or flipping) of the "direction" of the carbon atomic $p_z$ orbitals, playing the role of pseudospin at the corresponding sites, which is also the reason for the above (b) HOMO-LUMO alternations and coupling.

**(d)** and finally the resulting two aromaticity patterns (Clar-type for odd n, versus non-Clar for even n)[1], which are complementary to each other (the full rings of the one correspond to the empty rings of the other),  as determined by the NICS(1) magnetic aromaticity index[14-15] keep alternating with n, as a consequence of the above alternations/interchanges. For infinite graphene this can be described as a resonance state between the two aromaticity patterns. Since each aromaticity pattern with ν Clar sextets corresponds to $2^{\nu}$ resonances (of kekulé type valence structures) the electronic resonance in graphene corresponds to the order of $2^{\nu}$ , $\nu \rightarrow \infty$ resonances of this type giving rise to  a "robust electronic coherence".

The deeper reason for such alternations, as will be further explained below, is inversion symmetry "frustration", present in the molecular $D_{6h}$ symmetry group, which is the "dominant" symmetry (in the molecular picture); but missing in the "sublattice" symmetry group ($D_{3h}$), which is the "prevailing" symmetry in the band picture (see Figs. S2(b), S2(c), and S2(d)). Thus, inversion symmetry "frustration" (or competition) is equivalent to competition between bonding and banding, or molecular versus crystalline description. At infinite size (and level) these two descriptions should be clearly equivalent (by dynamic interchanges or fluctuations). Obviously, the "sublattice structure" is really meaningful for the infinite "crystal". For the finite models of Fig.1 (or Fig. S2) for which not translational invariance has been explicitly introduced, the difference between carbons at sublattices A or B comes into play gradually through the different "environment" (chirality in particular) at the two sites. Each layer addition in the PAHs of the main sequence can be topologically seen as a rotation by $60^{\circ}$ of their "principal" symmetry axis (as well as a substrate interchange at least in the outer layer), so that an odd layer number (odd n) corresponds to underlying trigonal symmetry of the arrangement of the frontier orbitals and of the corresponding aromaticity pattern, whereas an even layer number leads to hexagonal overall symmetry ($120^{\circ}$).On the other hand, as can be verified by Fig. S1, a $60^{\circ}$ rotation interchange sublattice sites, which reveals the fundamental interconnection between the number of aromaticity patterns and sublattice structures. Such interconnection can be used to define an "effective sublattice structure" for each PAH, all the way to infinity, based on the symmetry of the HOMO orbital, or on the type and/or symmetry of the aromaticity pattern. As is shown in Fig. 1, the filled rings of the PAH with shell number $n$ consist of the empty (core) rings of the PAH with shell number $n-1$, which plays the role of the "soft core" in the shell structure (see Fig. S2(a)), and the periphery (valence) annulene ring ($n$). This is a consequence of the coupling relations (S1), (S2), (or eqs. 12, 13 in ref. 1). Thus, the full sequence of the "full" and "empty" rings, together with the "full" structure and symmetry of HOMOs and LUMOs, can be generated easily in a straightforward way for each and every one of the PAHs of the "main sequence", using relations S1, S2 successively for n=2, 3, ….), revealing at the same time the electronic, aromatic, and structural "composition" of the given PAH in terms of the constituents benzene rings. This is obviously true for every n, and presumably, for graphene (n→∞). Fig. S3 illustrates that electronic properties, such as the HOMO-LUMO gap and the average number of π-electrons per ring converge to the correct limits (0 and 2, respectively). Thus, at the infinite limit of graphene the above alternating properties as n alternates between odd and even values must coincide and couple dynamically; until a final conduction-valence band inversion, associated with a dynamical breakdown of parity occurs, as in the many-body theory of graphene,[5] turning the molecular group from $D_{6h}$ to $D_{3h}$ with no inversion symmetry and identical HOMO LUMO orbitals of the same E" symmetry. This is because, as was mentioned above (and shown earlier[1]) there is a coupling between the HOMO and LUMO frontier orbitals not only of a given PAH (PAH[n]), but also of its near neighbors PAH[n-1] and PAH[n+1], (see relations S1, S2, S1', S2' in the supplementary information), which can be also verified by a careful inspection of the structure and symmetry of the HOMO and LUMO orbitals of the PAHs in Fig.1, which in fact constitute a convolution or multiple "reflections" of interwoven benzene's HOMO and LUMO orbitals. Thus, the interchange between (benzene-like) HOMO and LUMO orbitals, or the interchange between "full" and "empty" rings "propagates" as n increases, apparently all the way from n=1 (benzene) to n →∞ (graphene). Obviously, such interchange for finite size PAHs should be interpreted as "coupling" of the corresponding properties in the limit of n→∞ (infinite graphene), since the results for n and n+1 (or n-1 and n) should be identical. The special significance of these results for graphene, seen as a gigantic PAH (in the limit of $n \rightarrow \infty$) is obvious (see Table S1 and associated discussion in SI). This result (the touching or "coupling" of HOMO and LUMO orbitals, or of the valence and conduction bands), which is rooted in the bonding versus banding competition (or equivalently to the molecular aromatic properties of benzene coupled with the topological properties of the honeycomb lattice), leads to the Dirac's points (or Dirac cones in k-space band-structure), which can be easily recognized in the ("simulated") density of states (DOS) of Fig. 2, which is generated by a gaussian broadening of the energy levels of the PAH with n=7, of Fig. 1, with real HOMO-LUMO gap=1.12 eV

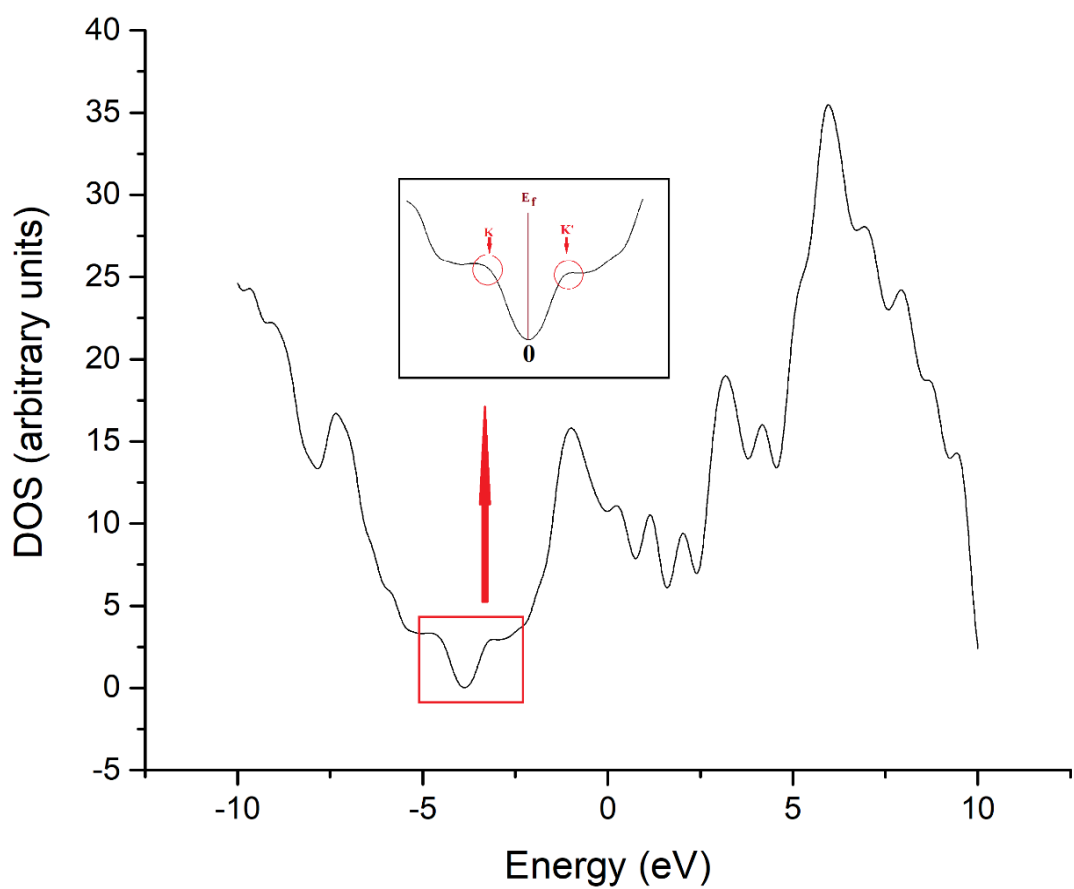

**FIGURE 2.** "Density of states" for the PAH with n=7, generated by a gaussian broadening of the energy levels by 0.30 eV. The inset emphasizes the Dirac (K, K') points around the Fermi level, $E_f$.

The coupling of HOMO and LUMO orbitals (which was proven here and was illustrated to be fully compatible with the many-body theory of graphene) is indirectly included in the (tight binding ) band structure of graphene, through the recognition of two different sublattices (with a $D_{3h}$ crystal symmetry) and the choice of the wavefunction $\Psi_k(R)$, for a given wavevector k as: $\Psi_k(R) = C_A \Psi^A_k + C_B \Psi^B_k$, where A and B are the two different sublattice sites[4]. As was illustrated earlier (see also Table S1), coupling of the HOMO and LUMO orbitals is associated with sublattice sites coupling, as well as coupling of the two aromaticity patterns. Therefore, we can say that the usual band structure, which corresponds to an "average picture" includes by construction as a fingerprint of such dynamical interchange(s) the Dirac's cone(s). Knowing the ultimate macroscopic reason (symmetry) for such coupling and/or interchange, the crucial question then is how this interchange takes place microscopically at the atomic scale? Physically, the shell structure and the resulting coupling between core and valence shell[1] frontier orbitals (see relations (S1), (S2)) could explain such interchange / coupling, but this is not the full story (or, at least, not the whole story). As we have already seen, topologically, the driving force is inversion symmetry invariance and the resulting *geometrical frustration* between sublattice and full (space group) symmetries (see Figs. S2(b), S2(c)). The same inversion symmetry frustration (or inversion symmetry breaking) is responsible for the "relativistic" (four-dimensional) behavior of graphene[3] (through the coupling of two two-dimensional HOMO, LUMO MOs), the dynamical breakdown of parity (and time reversal invariance)[4-5] etc.[5-8] To continue further such qualitative analysis, the results and properties in Table S1, describing the dynamical (coordinated) coupling of the two "phases" (or "stages") for all interrelated properties, should be connected by topological (geometrical) and physical aspects at the fundamental atomic level. To this end, consider

Fig. 3 which shows the elementary construction of the six $\pi$ MOs of benzene ($\pi_1$, $\pi_2$, …, $\pi_6$) in terms of the six constituent $p_z$ carbon atomic orbitals (AOs). As shown in Fig. 3, the lowest energy MO of $a_{2u}$ symmetry consists of six $p_z$ AOs on each corner of the ring, all aligned in the same way with similar phases in nearby atoms (thus overlapping constructively) without any nodes at all. Second lower in energy is the doubly degenerate $e_{1g}$ bonding MO characterized by one nodal plane through the atoms, for $\pi_2$ (on the left), or through the bonds, for $\pi_3$ (on the right). Higher in energy is the doubly degenerate $e_{2u}$ antibonding MO ($\pi_4$, $\pi_5$) characterized by two nodal planes: one through the atoms and one through the bonds. Finally, highest in energy is the $b_{2g}$ (or $\pi_6$) non-degenerate MO with three nodal planes through each bond. It is important to observe in Fig. 3 that for all odd parity (u) MOs, inversion through the center of the ring transforms (maps) a $p_z$ AO to an equivalent one with the same orientation (phase). For the even parity (g) MOs, the opposite is true; i.e. inversion of a $p_z$ AO through the center of the ring leads to a $p_z$ AO with opposite orientation (opposite phase), which is also verified in Fig. 3(b) for the corresponding $\pi$ MOs of benzene. Then, loosely speaking, we can say that u-symmetric MOs (and in particular HOMOs $e_{2u}$) reflect the molecular (hexagonal) symmetry, and "hexagonal" (CO) aromaticity pattern; whereas g-symmetric ($e_{1g}$) HOMO reflects crystalline (sublattice) symmetry and "trigonal" (CIRCO) aromaticity pattern. Moreover, besides benzene, since the HOMO and LUMO orbitals of the main sequence of PAHs (generated by the shell model) in fact consist of interwoven benzene HOMOs and LUMOs, this is also true for the larger PAHs' HOMOs and LUMOs with respect to their center, as can be clearly seen in Fig. 1. This property is extremely important for establishing a connection with the (topological) concept of the two sublattices of graphene since inversion through the origin (center) interchanges sublattices. This should be understood in connection with the "prescription" for interchanging u and g MOs (and therefore HOMO and LUMO) by "flipping" the "suitable" $p_z$ AOs, given at the bottom of Fig. 3. As can be seen in this figure, by flipping the $p_z$ AOs in positions 2, 4, and 6 in the opposite z direction (to -2, -4, -6 ) the $e_{2u}$ MOs automatically transform to $e_{1g}$ (or HOMO → LUMO, HOMO ← LUMO) and vice versa. Moreover, if we temporarily consider that each ring in Fig. 3 is part of a graphene lattice, it becomes immediately clear that the carbon atoms in the positions 2, 4, and 6, all belong to the same sublattice. Therefore, it can be concluded that the (dynamical) interchange of CO and CIRCO aromaticity patterns, or the interchange of "full" and "empty" rings, or even, equivalently, the swapping of HOMO LUMO MOs are fundamentally generated by the flipping of the $p_z$ AOs located in either one of the two sublattices.

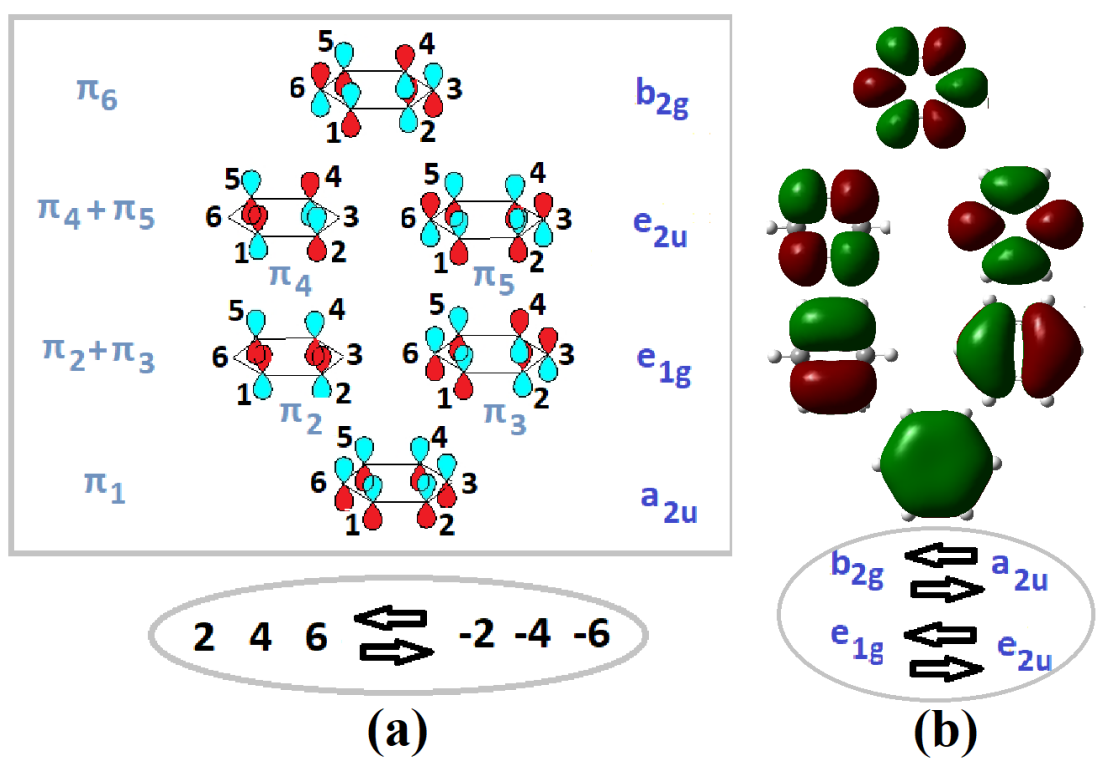

**(a)**                    **(b)**

**FIGURE 3**. The arrangement of the $p_z$ atomic orbitals of the carbon atoms in benzene (a) for the construction of the corresponding six π MOs (b). The oval scheme(s) at the bottom of the figure indicate the flipping of the $p_z$ AOs at the particular ring positions

and the resulting transformations of odd parity (u) MOs to even parity (g) MOs and vice versa.

The driving force is inversion symmetry, since inversion, as we have illustrated earlier, interchanges sublattices. Each sublattice separately, having planar trigonal ($D_{3h}$) symmetry, lacks a center of inversion, which the full $D_{6h}$ symmetry has, thus leading to a "staggering" "sublattice effect" or "staggering" $p_z$ orientation, reminiscent of the staggering magnetic spin (or pseudospin[3]) effect. From a different, but equivalent, point of view such effect can be seen as an attempt of the system to respond to the different chemical environment of (the even-numbered) carbon atoms 2, 4, 6, in comparison to the (odd-numbered) atoms 1, 3, 5 (which are otherwise chemically identical), when the ring is part of the graphene lattice. This is because in the u-representations, as was seen earlier, the $p_z$ orbitals in the two sublattices (i.e. in the positions 1,3,5 and 2,4,6) appear fully equivalent (with the same phase and orientation); whereas in the g- representations the $p_z$ orbitals "correctly" have opposite orientation (and phase) in such (opposite sublattice) sites, allowing or reflecting the distinction of different chemical environments (e.g. chirality). Obviously, the same is true for the resulting MOs, as can be verified in Figs. 1 and 3. Thus, the even representations (g) reflect the lattice (sublattice) symmetry as well as the symmetry of the CIRCO aromaticity pattern (i.e. trigonal, $D_{3h}$); whereas the odd (u) representations reflect the symmetry of the molecular (full) group and the symmetry of the corresponding CO aromaticity pattern (i.e. hexagonal, $D_{6h}$). Therefore, the competition or interchange and (final) coupling between odd-even, HOMO-LUMO orbitals, as well as all the resulting interchanging and couplings (illustrated in Table S2) are the results of the tendency to balance the symmetries of the $D_{3h}$ (sub)lattice with the overall $D_{6h}$ molecular symmetry. This is accomplished in infinite graphene at the Dirac points (at the $D_{3h}$ symmetric K points in k-space), where the spatial and crystalline symmetries meet. Note that the key to this balance is inversion symmetry, which is satisfied in the molecular symmetry group ($D_{6h}$) but missing in the (sub)lattice group ($D_{3h}$). One could also see microscopically this "sublattice effect" (through the flipping of the $p_z$ AOs) as an analog of a ferromagnetic-antiferromagnetic transition (compare the $a_{2u}$ and $b_{2g}$ or the $\pi_1$ and $\pi_6$ MOs in Fig. 3). Alternatively, one can see such $p_z$ AO flipping resulting in a HOMO, LUMO orbital interchange, or "crossing" as a phase change both, of the constituent AOs, and of the HOMO / LUMO wavefunction by a factor of $\pi$ (u →g, or u←g), "protected" by inversion symmetry, in analogy to Berry phases (in k-space) in topological insulators. However, no topological edge states have been explicitly observed in the HOMO and LUMO orbitals of the hexagonal samples we have examined in the present study, although the shell model concept by definition supports valence and conduction frontier MOs "concentrated" in or "dominated" by contributions in the region of valence and conduction shells respectively. This, however, is a natural and expected result (contrary to edge-localized and "polarized" states in rectangular NGRs and GNRs, which have a totally different origin). Edge (or periphery) "localized states" should be clearly visible for larger PAHs with surface-dominated HOMOs and LUMOs, or even for smaller PAHs at the "suitable" (relatively large) isovalue. It is interesting to observe that in the PAHs in Fig.1, with the possible exception of the first two (benzene and coronene) which could be seen differently, all edges are of zigzag type, which are usually "demonized" for spin polarized states and other "peculiarities", *vide infra.* Yet, this is not true, or at least this is not generally true as in the present case involving hexagonal symmetry. Zigzag edges are "normal" or "well-behaving" edges, at least as normal as armchair edges are (for hexagonal topology). It is clearly shown in Fig. S4 that the same general rules which relate the parity of the shell number $n$ or $n_{eff}$ (see ref. 1) with the aromaticity pattern, and the symmetry of the HOMO, LUMO orbitals are invariably valid for armchair PAHs as well. Nevertheless, since armchair or zigzag edges are indispensable part of the topology of the samples (with armchair edges symmetrical about the $C_2''$ or x-axis "through the bonds" in Fig. S1, while zigzag edges are symmetrical about the $C_2'$, y-axis "through the atoms) there would always be an indirect (but vital) link of molecular (or crystalline) symmetry with the associated edge types, and the corresponding aromaticity pattern(s) of a given PAH or NGR. Thus, in general the properties of PAHs or NGRs at the edges should be separately described for armchair and zigzag edges, whereas some combinations of "peculiar" topologies and edge types could be "problematic". Physically, the absence of edge states in hexagonal PAHs can be understood in terms of the core-surface (edge) coupling[1] described by relations (S1) and (S2), illustrating that both HOMO and LUMO orbitals are delocalized in the entire PAH (both in core and surface shell), but it is expected (especially for larger PAHs) that frontier MOs will be dominated by topologically frontier (*e.g.* "valence" and "conduction") shell states. The lack of "polarized" edge states in hexagonal PAHs is clearly connected to the special relation of the overall $D_{6h}$ molecular symmetry group and its full $D_{3h}$ subgroup, which describes the (sub)lattice symmetry; while both symmetry groups are two of the three unique symmetry groups (the third been the tetragonal $D_{4h}$) which allow regular tessellations. Clearly, for PAHs or NGRs of non-hexagonal symmetry, such as GNR's of $D_{2h}$ symmetry this is no longer true. Rectangular NGRs of D2h symmetry, in particular, always include both zigzag and armchair edges at right angle to each other. Therefore, if topological gapless edge states are going to appear in the HOMOs and the LUMOs (as a result of "sublattice frustration") at all, rectangular NGRs or GNRs, contrary to hexagonal ones, should be the best candidates for this. The results below fully confirm such expectation.

**3. Rectangular NGRs and GNRs.** Obviously, for non-hexagonal NGRs, although the region of the central hexagonal core could be still somehow defined, the surface shell is not always meaningful or unique (and, obviously, non-isotropic). Furthermore, due to symmetry lowering the degeneracy of the 2D frontier MOs $e_{1g}$ and $e_{2u}$ ($\pi_2$, $\pi_3$ and $\pi_4$, $\pi_5$ in Fig. 2) will be lifted and each 2D MO will be reduced to two 1D MOs of proper symmetry; while the coupling relations (S1), (S2) for non-hexagonal samples are no longer true or meaningful (at least as they are, without modification). Therefore, a decoupling of core and surface frontier MOs could be possible and could be expected, leading to surface-localized HOMOs and/or LUMOs (at the zigzag edges, for reasons which will become clear below). It should be stressed at this point that although the shell-model was derived for hexagonal PAHs, the shell structure is a much more general effect, unfolding the core-surface interrelation and manifested through the periodicity of aromaticity patterns for both hexagonal and rectangular NGRs as more new surface layers are added. Such periodicity in the aromaticity patterns was recognized[2] well before the shell model was established and the shell structure was revealed[1] or understood. This is why aromaticity is a good starting (and ending…) point for bridging the physics and chemistry of graphene (through bridging benzene and graphene) and unifying the shell-structure description of hexagonal and rectangular NGRs (and GNRs). The geometrical/topological and (therefore) aromatic relationship(s) between rectangular and hexagonal NGRs are illustrated in Fig. 4, in which the rectangular NGRs consisting of Z zigzag rings and A armchair rings at their edges are labeled as ZxA. Figure 4(a) shows ZxA rectangular NGRs or GNRs of $D_{2h}$ symmetry in order of increasing Z (Z=1-7, and A=Z), together with inscribed hexagonal NGRs, with shell number *n* (shown below the NGRs) demonstrating the shell structure. Figure 4(b) shows the two key hexagonal PAHs: coronene (CO), with *n*=2, and circumcoronene (CIRCO), with *n*=3, together with their characteristic aromaticity patterns (CO and CIRCO). It is customary, based on Clar's analysis[10-11] for AGNRs, to call the CIRCO pattern *Clar* pattern, and the coronene (CO) *Kekulé* pattern, reserving the term *incomplete Clar* for the mixed aromaticity pattern of the Z=3n-1 AGNRs[11]. The same patterns have been termed alternatively on the basis of the number of Clar's sextets (C) in the unit cell, as 1C, 2C , and multiple C, nC, respectively.[10] Here, without resorting to empirical Clar's rules (which are indirectly included in the shell model anyway[1]) we keep the CO and CIRCO terms for the description of the key patterns from the corresponding prototype PAHs of the main sequence) to remind us the deeper origin and meaning of the terms from the shell structure (CIRCO pattern has aromatic central ring compatible with odd shell number n; CO pattern has non-aromatic central ring and even n[1]). It is clear, as was expected, that for rectangular NGRs we have a three membered periodicity[2], corresponding to the two-member periodicity of the hexagonal ones. Then, for a given A (A=10 in Figs. 4(c), 4(d)) the rectangular NGRs can be classified with respect to their aromaticity pattern (and all associated electronic and cohesive properties[1-2]) in three categories according to the number Z of zigzag rings as Z=3n-1, Z= 3n , Z=3n+1 , where n =1, 2, … is an integer (not to be confused with the shell number *n* for hexagonal PAHs). Fig. 4(c) for n=1, and Fig. 4(d) for n=2, represent the aromaticity patterns of three characteristic GNRs, or more precisely AGNRs. It is clear from Fig.4 that AGNRs of a given width, W, have the same aromaticity pattern independent of their length, as can be also verified in Fig. S5. Notice also that the sublattice symmetry requirement is fulfilled at the armchair edges (neighbouring carbon atoms belong to different sublattices). Therefore, AGNRs are commonly classified by their width W which is usually defined by the number of carbon atoms across the zigzag edges[14-16], which in terms of Z is given by W=2Z+1; whereas their length (in the very rare cases in which it is considered[12]) is simply given in terms of the number of carbon atoms along the armchair edges, L (L=2A).

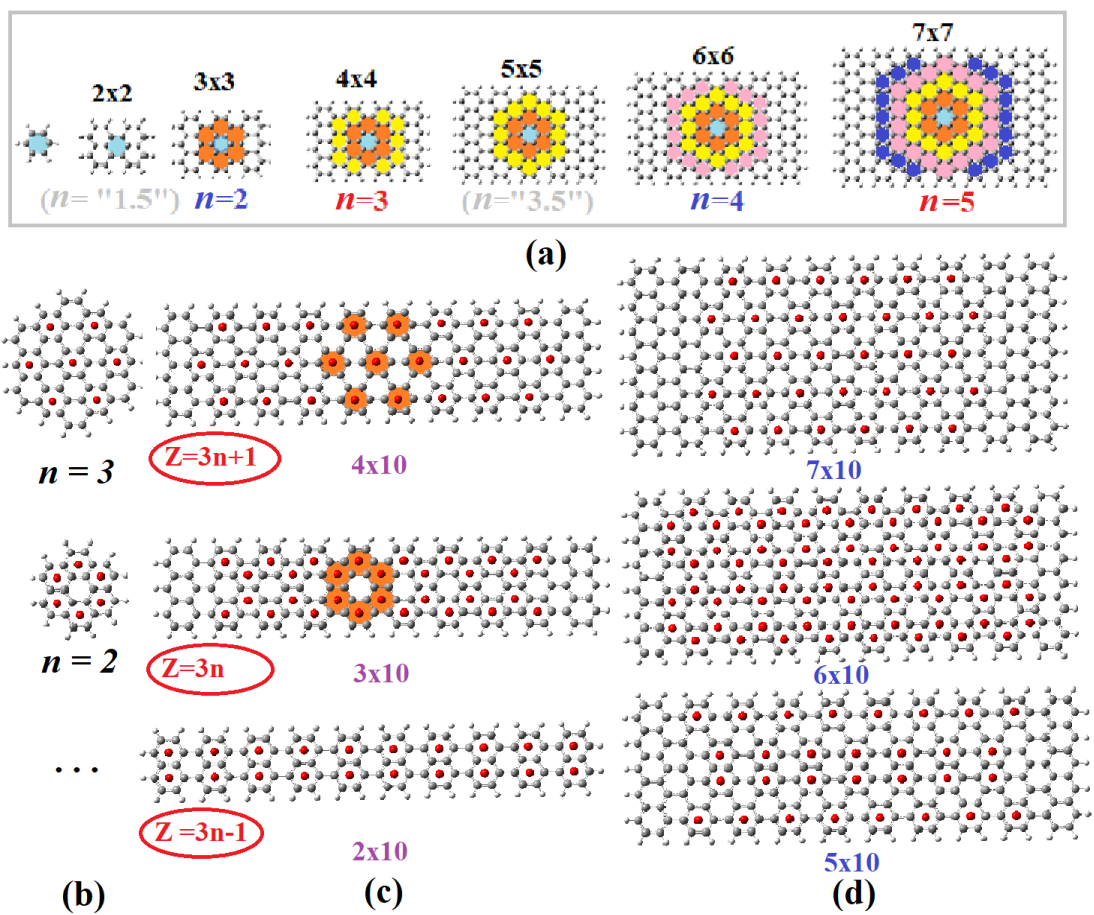

**FIGURE 4. (a):** Seven ZxA rectangular NGRs, with Z=A=1-7 with inscribed schematic hexagonal shell structure. The corresponding hexagonal shell numbers *n* are shown below the structures. **(b)**, **(c)**, and **(d):** Comparison of hexagonal (a), and rectangular (b)-(c), aromaticity patterns for GNRs with Z =3n-1, 3n, 3n+1; where n=1 (c), n=2 (d), and A=10. Aromatic ("full") rings are shown with red (on line) dots in their centers (see text).

Thus, the AGNRs in Fig. 4(c) from bottom to top are characterized by W=5, 7, and 9, respectively and L=20; and those of Fig. 4(d) by W=11, 13, 15 (L=20). As can be seen in Figs. 3 (b, c, d), the Z=3n+1 AGNRs  4x4, 7x7, etc. (or W=9,15, etc.) correspond to odd shell numbers and are characterized by Clar-type CIRCO aromaticity patterns; whereas the Z=3n type AGNRs (3x3, 6x6 etc., or W=7, 13, etc.) correspond to even shell numbers and are characterized by CO aromaticity patterns. Both of these aromatic types correspond to the peaks of the shell model ("shell closures")[1] and are therefore expected to have (and they indeed have) large HOMO-LUMO gaps or bandgaps. The Z=3n-1 type of AGNRs, on the other hand, have contributions of two different (successive) shell numbers, shown by non-integer shell numbers in Fig. 4(a) and cannot be assigned a unique shell number. As a result, their aromatic and electronic properties cannot come to one-to-one correspondence with those of the shell model. Therefore, the Z=3n-1 AGNRs correspond to mixt non-coherent aromaticity patterns (see bottom of Fig. 4(d)) and have small (or near zero) bandgaps. Thus, the present approach provides a full and attractive explanation for the bandgap properties of AGNRs, which although well-known[10-13, 18-20], are not fully understood. Further understanding is gained by careful examination of the HOMO, LUMO orbitals, and their energy separation (HOMO-LUMO gaps), the structure and symmetry properties of which are closely connected with aromaticity and aromaticity patterns, which, in turn, are directly connected with topological characteristics and symmetry properties. Such symmetry properties (and in particular inversion symmetry) of the full molecular group ("space group") versus the (sub)lattice symmetry and structure are largely responsible for the appearance of gapless topological edge states. From the MO-symmetry point of view the changing (lowering) of $D_{6h}$ symmetry involves the reduction of the (2D) $D_{6h}$ MOs to (1D) $D_{2h}$ ones, through the compatibility relations of the corresponding groups[21] (see also scheme S1, and section #S7.2 in the supplementary information), which show that $e_{1g}$ and $e_{2u}$ $\pi$ MOs would be reduced as:

$e_{1g} \rightarrow b_{1g}+b_{2g}$     and

$$e_{2u} \rightarrow a_u + b_{3u} .$$

The new $D_{2h}$ MOs can be distinguished in those which are symmetrical about the y-axis (Fig.5), as the $b_{3u}$ and $b_{1g}$ in Fig.5b; and those which are antisymmetric with respect to the same axis, as the $a_u$ and $b_{2g}$ MOs in Fig.5b. The antisymmetric MOs reflect (are compatible with) the sublattice symmetry, demanding opposite orientations (phases, or colors) at the two zigzag ends. The symmetrical (about the y-axis) MOs, on the other hand, reflect the full molecular ($D_{2h}$) symmetry, demanding identical orientations (phases and colors) at the two zigzag ends. In addition, the new $D_{2h}$ MOs could be further distinguished in core-like and surface-like, since the coupling relations (eqs. S1, S2) are no longer valid. This is illustrated in Fig. 5 for the Clar-type CIRCO ($n$=3) and the corresponding 4x20 (or W=7, L=40) AGNR, belonging to the Z=3n+1 category.

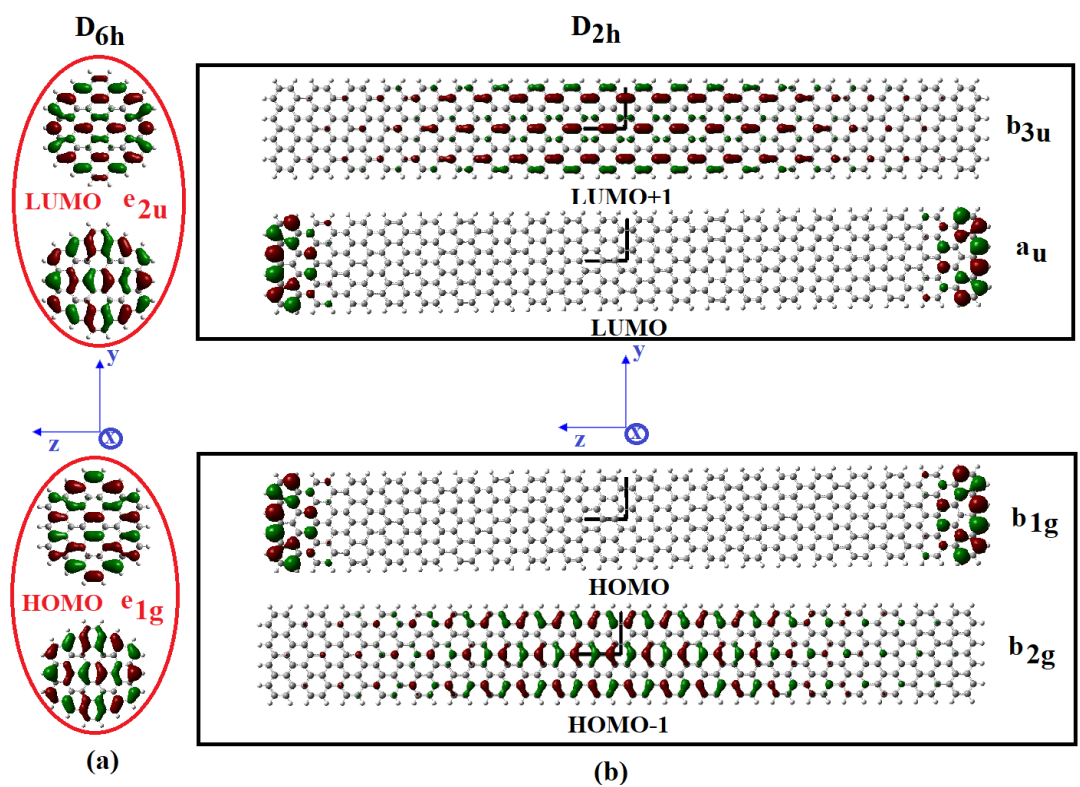

**FIGURE 5**. Demonstration of gapless topological edge states. Comparison of the frontier orbitals of (a) hexagonal NGRs (for $n$=3); and (b) the corresponding rectangular NGR (AGNRs), with Z=4, A=20, (4x20), or (W=7, L=40).

Observe that:

(1st) the parity and energy ordering of the occupied and unoccupied MOs is preserved, so that the hexagonal (HOMO) orbital corresponds to the rectangular (HOMO-1)+(HOMO) orbitals, and similarly the "companion" (LUMO) to the rectangular (LUMO) and (LUMO+1) orbitals. This is not true for the Z=3n-1 AGNRs which have a mixt aromaticity pattern and, subsequently, mixt HOMO-LUMO components.

(2nd) Both new ($d_{2h}$) HOMO and LUMO orbitals are topologically "decoupled" edge ("surface") MOs. Such edge states have been predicted earlier by DFT calculations[2,12-13] and observed experimentally by scanning tunneling spectroscopy (STS),[19-20] but their nature and origin was not fully known or understood till now.

(3rd) In the occupied even-parity MOs the "surface" $b_{1g}$ MO, which is symmetrical about the y-axis is higher energetically, whereas in the unoccupied odd-parity MOs the antisymmetric (about the y-axis) $a_u$ "surface" state is energetically lower, leading to gapless edge states as n →∞. Physically, as is further illustrated in Fig. 6, this happens because for the occupied MOs the sublattice symmetry is the important factor determining the energy order; whereas for the unoccupied MOs the overall molecular symmetry is responsible for the energetical ordering. More specifically, from the two even-parity modes ($b_{1g}$ and $b_{2g}$) the $b_{1g}$ representation is incompatible with sublattice symmetry, since, as was shown earlier, is antisymmetric with respect to rotation about the y- axis or the vertical plane xy. This produces carbon atoms with $p_z$ AOs (and therefore MOs) with identical orientation (phase) at opposite zigzag edges, which for even parity (g) representations demands carbon atoms of the same sublattice. On the other hand, the $b_{2g}$ representation being symmetric (see character table in the scheme S1) with respect to y- rotations, is compatible with sublattice requirements (with no sublattice frustration at the central ("bulk") region, as is shown in Fig. 6(a) top) and therefore represents a "bulk" MO of lower energy. For similar (but not identical) reasons the $a_u$ representation from the unoccupied odd-parity (u) MOs, which is symmetrical with respect to $180^o$ rotation about the y-axis, corresponds to a surface (edge) MO due to sublattice frustration (at the central region, as is illustrated in Fig. 6(a)) since it describes carbon atoms at opposite zigzag edges with opposite phases, whereas for u representations C atoms at opposite sublattice must have $p_z$ AOs (and MOs) with identical orientation and phase. However, this surface-like $a_u$ unoccupied MO, satisfying the overall molecular group symmetry, would be lower energetically compared to the bulk-like $b_{3u}$ MO, since for unoccupied MOs (contrary to occupied ones) the (higher) molecular symmetry determines the energetical ordering. Such difference between occupied and unoccupied states is related with the electron-electron Coulomb interaction, which is operative between electrons in occupied MOs, as will be seen further below. The important conclusion of this discussion thus far is that the lowering of symmetry from hexagonal ($D_{6h}$) to rectangular ($D_{2h}$) pushes the resulting occupied edge HOMOs higher in energy, while the unoccupied edge LUMOs are pushed lower in energy. As a result, the HOMO-LUMO gap between edge ("surface") states is progressively diminishing ($\Delta E=0.02$ eV for the edge states in Fig. 5), approaching zero at infinity. Thus, these states are essentially gapless (protected by inversion symmetry), as would be expected for topological edge states. These edge states are clearly nonbonding, and their full characteristics are primarily determined by geometrical and topological factors and not so much by the Chemistry of the sample; and additionally, due to their small range of localization they do not contribute significantly to transitions from occupied to unoccupied states. As a result, transitions from occupied (fully) edge-localized states to (fully) edge localized unoccupied states should be difficult to observe experimentally, contrary to "bulk"-to-"bulk" or even "bulk"-to-"edge" transitions, leading to incorrect identification to measured (STS) peaks[12].

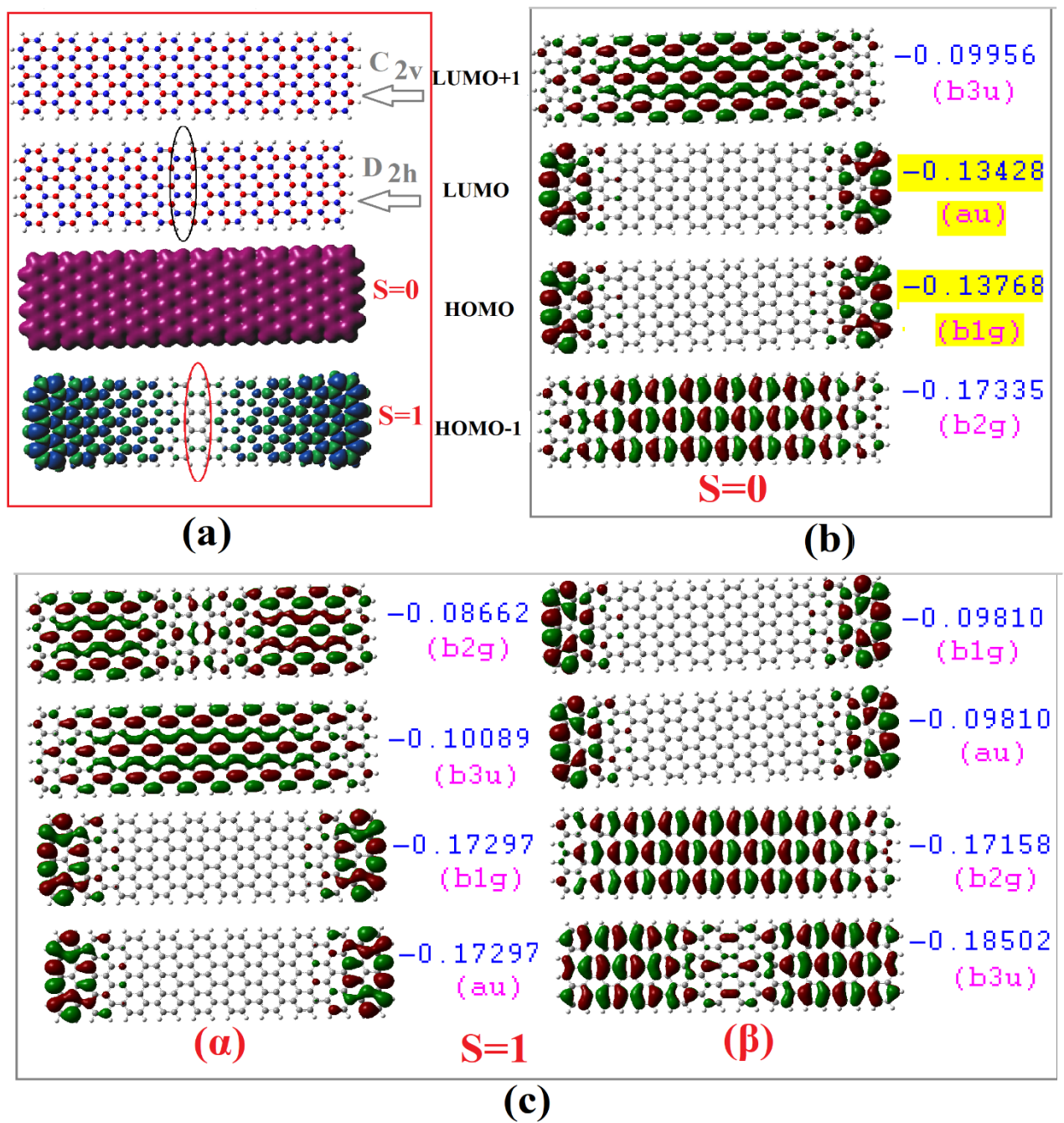

**FIGURE 6**. Rectangular nanographenes of $D_{2h}$ symmetry: **(a)** The geometrical structure of the two sublattices A and B, shown with red and blue (on line) spheres respectively; below is the same sublattice structure required by the $D_{2h}$ symmetry group starting from the zigzag edges. Third from the top is the charge density. Lowest in the bottom (left) is the spin density of a "hypothetical" triplet state (see text), shown in blue and green color (on line) for the majority and minority spins, respectively. The region of "geometrical frustration" is shown with vertical ellipses. **(b)** The frontier orbitals HOMO-1, HOMO, LUMO, and LUMO+1 for the singlet ground state, together with their energies (in a.u.) and symmetry (in parenthesis) in their right-hand side. **(c)** The corresponding singly occupied (frontier) molecular orbitals (SOMOs) and information as in (b), for the (α) and (β) spins of the triplet state, obtained by unrestricted DFT.

Thus, the effective HOMO-LUMO gap (with efficient transitions, and relatively large oscillation strengths) should be considered the energetical distance between the bulk-like HOMO-1 and the corresponding LUMO+1 MOs, i.e. the "bulk gap".[12-13] By comparing to Figs. 5(a) and 3(a), it becomes clear these orbitals should indeed be real HOMO and LUMO, since they both originate from the $\pi_2$ and $\pi_4$ HOMO and LUMO orbitals, respectively, of the hexagonal "parents", compatible with $D_{2h}$ symmetry (the $\pi3$ and $\pi5$, which are incompatible with rectangular symmetry are reduced to the practically isoenergetic edge states). In this respect it is very interesting to compare in Fig. 4(a) the real (standard) HOMO-LUMO gap of the "parent" *n*=3 hexagonal PAH, CIRCO ($C_{54}H_{18}$), with the effective HOMO-LUMO gap one of the rectangular 4x4 offspring; The former is 2.8 eV, while the later 2.9 eV in very good agreement. Moreover, such ($\pi_2$-$\pi_4$) values are very close to the experimentally measured HOMO-LUMO gap[12-16] (or the "optical gap", determined by transitions with "relatively high" oscillator strength, which are dominated by HOMO* to LUMO* excitations),[12-13] as well as many-body calculations[24] for AGNRs similar to the ones studied here. It must be emphasized at this point that all these features (1st, 2nd, 3rd) are based on the shell model (which is founded on physical and topological principles[1]), and (inversion) symmetry requirements (and conflicts) of the full molecular group ("space group") and the sublattice structure. This is explained better in the upper part of Fig. 6(a), where the role of the sublattice symmetry is illustrated in more detail. As is clearly seen, the sublattice symmetry is $C_{2v}$ with no center of inversion, whereas the full molecular symmetry group (or "space group") is $D_{2h}$ with inversion symmetry. If one tries to establish a sublattice structure keeping the $D_{2h}$ symmetry starting from the zigzag edges (since the armchair edges already fulfil sublattice requirements), one will end up with a region of sublattice frustration, shown with vertical ellipse in Fig 6(a) second from the top, exactly in the middle of the AGNR around the (inversion) center. This is because the rectangular molecular symmetry demands that carbon atoms in the two opposite zigzag edges should belong to the same sublattice (be "identical"); whereas sublattice symmetry dictates exactly the opposite (opposite sublattice). The result of such sublattice frustration (at the central region of the AGNR), as was shown earlier, is the formation of zigzag edge ("surface-like") HOMO and LUMO orbitals with very small (diminishingly small) HOMO-LUMO gap (of the order of $10^{-2}$ eV for the AGNRs of Figs. 4 and 5), which could be indicative of diradical behavior. It is interesting to note that such diradical behavior could be suggested by the *alternant hydrocarbon*[22] nature of graphene samples and graphene itself (as a bipartite lattice), since Hückel's theory predicts[22] a diradical ground state for neutral alternant hydrocarbons for which the number of "starred" (A sublattice) sites is non-equal to the number of "unstarred" (B sublattice) sites. However, this is not the case here, since as can be clearly seen in Fig. 6(a), the number of sublattice sites is the same but there is only sublattice symmetry mismatch at the central region of rings. Real diradical behavior (involving real spins) can in fact be encountered for other possible topological combinations of symmetry and edge type(s). This will be examined in a forthcoming publication. Interestingly enough, Hückel's theory also predicts for alternant hydrocarbons[22] (bipartite lattices) that energy levels are symmetrically arranged about zero (Fermi level) with the same orbital coefficients for occupied and unoccupied MOs except for alternating plus and minus signs; features that could be correlated to the form of conduction and valence bands of graphene and the postulated coupling of HOMO-LUMO orbitals. The unrestricted DFT calculations indeed suggest a triplet ground-state, which is really lower in energy than the singlet state (by about 1eV for the AGNR of Fig. 6). However, there are also opposite conflicting reports in the literature about this (see for example ref. 23), suggesting that this is only an artifact of the one-body (mean field) approach for a highly correlated system, which is largely true. But this is not the full story, because the unrestricted DFT results could contain and convey a lot of true new and transparent important physical (and other, e.g. topological) information. As is revealed in (the bottom of) Fig. 6(a), the lowest (unrestricted DFT) energy state is not just a usual spin triplet, but a pseudospin triplet, related with sublattice symmetry frustration at the central region. To see this better consider that the sublattice degree of freedom (A/B) is visualized as pseudospin (up/down) in each carbon site. Then the pseudospin density will be expected to show a pseudospin-glass region (with zero average pseudospin) right in the middle of the AGNR (where the sublattice frustration is maximized) together with an antiferromagnetic structure in the rest of the AGNR, where the sublattice and molecular ("space-group") symmetries can be locally compromised. This is indeed the case, as is emphatically illustrated at the bottom of Fig. 6(a). The energy gain in the pseudo-triplet state (within the present DFT results) can be further clarified by comparing the orbital energies in Figs. 6(b) and 6(c), showing lower orbital energies for the triplet state. This can be qualitatively rationalized in terms of a dropping in the (positive) electron-electron repulsion for electrons with the same pseudospin (electrons in the same sublattice) which are effectively kept on the average at a larger distance due to Pauli principle. Thus, the sublattice structure (which, as was illustrated earlier, is crucial for the energetical ordering of the occupied states) is indirectly connected with the long-range electron-electron Coulomb interactions. The importance of the long-range Coulomb interactions has been very well established in the many-body theory of graphene[7]. It has been shown[5-7] in the many-body theory of Dirac fermions describing the graphene layer, that a critical behavior associated with a dynamical breakdown of the parity (and time reversal) invariance can occur at sufficiently large strength of the long-range Coulomb interaction, through a mass term breaking parity (so-called Haldane mass). Thus, it seems that all (or at least, most of) the pieces of the graphene puzzle have been successfully fitted together. Yet, there are still more new results hidden in Fig. 6. Closer inspection of Fig. 6(c) further reveals that the pseudospin triplet ground state significantly ($\approx 100\%$) improves the postulated (here) coupling of the g-u MOs both in HOMO (AHOMO-1, AHOMO) and LUMO (BLUMO, BLUMO+1) SOMOs, through the ($\approx 100\%$) energetical overlap of the g – u states. Moreover, such coupling works equally well ($\approx 100\%$) separately between edge states as well as between core states, as would be expected, since they both come from the decoupling of the "original" hexagonal "core + surface" $e_{1g}$ and $e_{2u}$ MOs. Finally, as can be seen from Fig. 5(b) the "effective" (or "core") HOMO-LUMO gap[12] for the AGNR of the figure (with W=9, L=20) comes out to be 2.0 eV in excellent agreement with the 2 eV value for the energy gap obtained by many-body theory (using the Green's function method in the GW approximation).[24] It should be emphasized that in the usual electronic structure calculations of graphene using the common periodic boundary conditions in k-space, no topological (or other) edge states can appear for graphene nor for AGNRs of infinite length. Yet for finite size AGNRs, as the atomically precise AGNRs, edge states exist and have been verified experimentally[19-20]; although transitions between edge-localized states (in contrast to transitions between "core" or between "core" – "edge" states), are not easily detectable by experiment due to their low spectral weight[12]. It is therefore suggested here that the "real" or "measured" band gaps

Should be given by the "effective" HOMO and LUMO orbitals[12], which in the present case are the HOMO-1 and LUMO+1 MOs. With this interpretation there is excellent agreement of the calculated bandgaps and the experimental measurements (see for instance ref. 12)

**4. Conclusions**. The most fundamental and concise conclusion of this work should be that all exotic properties of graphene stem from the competition between the molecular and crystalline nature of graphene, which is shown to essentially consist of interwoven benzene molecules. Although this conclusion could be considered "natural" or even "trivial", it has not in fact been illustrated or "spelled out" elsewhere before, according to the present author's knowledge. Alternatively, graphene can be also seen as a macroscopic (crystalline) manifestation of (molecular) aromaticity (of benzene), which appears to be a topological property in the end, so that all "exotic" properties of graphene can be considered as a coherent combination of the aromatic properties of benzene and the topological properties of the honeycomb lattice. More specifically, the Dirac points, where the valence and conduction bands meet, are due to the topological competition between full $D_{6h}$ ("molecular") and crystalline $D_{3h}$ ("sublattice") symmetry, the latter been also the k-point group symmetry at the K and K' points. The same competition, which is essentially a "bond-versus-band" or "molecular-versus-crystalline" competition, leads to the interchange and final coincidence ("touching") of the symmetry (and energies) of the doubly degenerate HOMO-LUMO orbitals (at both $D_{6h}$ and $D_{3h}$ "high symmetries" with $e_{1g} \rightleftharpoons e_{2u}$ or $e_{2u} \rightleftharpoons e_{1g}$, and e''$\rightleftharpoons$e'' interchanges, respectively). Note that the in final (n $\rightarrow\infty$) "crystalline" $D_{3h}$ geometry both HOMO and LUMO have the same e'' symmetry (with no inversion). Equivalently, we have an interchange and final coincidence of CO and CIRCO aromaticity patterns (as well as "sublattice type" interchange) in the hexagonal PAHs of the main sequence, as more and more shells are added (up to infinity). Such interchange or coupling of the 2D HOMO and LUMO hexagonal orbitals also leads to a 4D representation and, eventually, to 4D Dirac-like "equation of state". The lack of inversion center in the $D_{3h}$ sublattice symmetry is also responsible for the conical form of the bands at Dirac points (in k-space), as well as the electron-hole symmetry in the half-filled $\pi$-band). Both are related to 1st) the special symmetry property between occupied and empty orbitals around the Fermi level ($e_{occ}$ $\rightarrow$ $-e_{unocc}$) of *alternant hydrocarbons*[22] (fully compatible with the sublattice structure), such as the PAHs of the main sequence, which is property and fully transferred to the corresponding band energy levels of the bipartite graphene lattice. 2nd ) As one moves away from the K point(s) at the edges of the Brillouin zone the symmetry of 2D E'' representation (in k-space) is reduced to two 1D representations, and the degeneracy is lifted even at first order of $\vec{k} \cdot \vec{p}$ perturbation theory, due to the lack of inversion symmetry in the $D_{3h}$ lattice of graphene (and the $D_{3h}$ k-space symmetry at the K points). The same property is also responsible for the electron-hole symmetry. It is interesting to observe that the underlying bipartite T.B. Hamiltonian obeys the $e_{occ}$ $\rightarrow$ $-e_{unocc}$ symmetry but only approximately when only first neighbor interactions are taken into account. However, in the present approach of graphene through the main sequence PAHs (within the full molecular orbital DFT theory, a slightly modified (generalized) similar symmetry still holds and is exact for the doubly degenerate HOMO and LUMO orbitals; in the sense that inversion operation (which is responsible for all interchanges we have witnessed so far) interchanges HOMO and LUMO orbitals (or the corresponding valence-conduction bands) with opposite symmetries (u $\rightarrow$g or g$\rightarrow$u), and with energies pretty much symmetrical (see for instance the density of states in Fig. 2, or Fig. S3(a). Another new emerging key-point here is that K points in rectangular graphene samples, symmetrically arranged around the Fermi level, appear as (correspond to) topological edge states at the zigzag edges which are essentially gapless (for large sizes), as a result of topological constrains related with (inversion) symmetry competition. The formation of such pairs of occupied/non-occupied states with marginal energy separation is due to the fact that the energetical ordering of occupied states is determined by the sublattice (crystalline) symmetry (which effectively minimizes Coulomb repulsion), in contrast to the non-occupied states where the energetical ordering is determined by the overall molecular symmetry, so that the occupied surface state (violating sublattice symmetry) would be energetically highest, whereas the non-occupied one (violating sublattice symmetry, but satisfying molecular symmetry) would be lowest. Thus, both HOMO and LUMO orbitals in AGNRs correspond exactly to such states, which, contrary to prevailing believes are not spin, but pseudospin polarized, depending on the mutual orientation of the atomic $p_z$ orbitals of carbon atoms at different sublattice sites. Also, contrary to prevailing views, if one considers these edge states as the "real" gap-determining HOMO and LUMO orbitals as is usually done[12], then AGNRs (similarly and in full correspondence to graphene) would be of essentially zero gap, yielding the same gap properties as graphene in terms of size, independently of the dimensionality of the samples (1D or 2D), as it should. This would be clearly in sharp contrast to the theoretical predictions[24] (which however do not include edge states in view of the periodic boundary conditions), and the experimental measurements.[18-20] If, on the other hand, one neglects these non-bonding topological edge states the remaining "effective" HOMO-LUMO gaps are obviously non zero, with very good agreement to experiment[12], and in part with the theoretical predictions. Such effective HOMO-LUMO orbitals, in contrast to the "ordinary" HOMO and LUMO orbitals have the same well-defined properties (symmetry) for each one of the three categories of AGNRs. Moreover, the present methodology has also led to the similarity of benzene and graphene and the conclusion that aromaticity is the key common property of benzene and graphene, which is actually (in all respects) a "super benzene", and similarly to benzene could be also considered as "super aromatic". Like benzene, graphene is a resonance structure (resonating between two, CO and CIRCO, aromaticity patterns), in line with the initial suggestion of Pauling.[2, 25] In the CIRCO pattern, corresponding to odd (hexagonal) shell number and even parity HOMO, the aromatic rings are arranged along the $C_2'$ axis of the central benzene ring ("trough the atoms"), see Fig. S1; whereas in the CO pattern, corresponding to even shell number and odd parity HOMO, the aromatic rings are arranged along the $C_2''$ axis ("through the bonds"); and this is independent of armchair or zigzag edges (see Fig. S4). Thus, graphene ("like benzene") should be considered as a prototypical aromatic crystal because it is: 1) a resonance structure, 2) planar, 3) stable, and in addition, (apparently) bottleable;[16-17] fulfilling even the most stringent criteria for aromaticity[16]. It should be emphasized that these ideas and conclusions were obtained using the simplest and most transparent way, based on the principles and conclusions of the shell model,[1] coupled with geometrical and topological arguments, connected with inversion, which is of crucial importance for the "balance" between sublattice(crystalline) and space group (molecular) symmetries, taking into account also that inversion interchanges sublattices. In conclusion, it has been demonstrated that the molecular nature of graphene is at least of equal, in not of higher, importance with the crystalline one, and the bridging of the two is not just a computational approach (as the title of the present work indicates), but a real and active physicochemical process which fully characterizes and generates all "peculiar" properties of graphene. Thus, the present approach could be proven very useful and innovative, providing fruitful guidelines for further studies not only for graphene, but for other 2D "crystals", and other graphene-based materials, such as GNRs and NGRs, as well.

**5. Methods**. The theoretical and computational framework of this work, which includes a multitude of systematic and interconnected (one-body) DFT calculations on PAHs, NGRs and GNRs (AGNRs) of given symmetry, has been discussed earlier,[1-2] together with the technical details. In the present work the "main sequence" of hexagonal PAHs, which defines the shell structure, has been expanded and extended to include analogous rectangular NGRs and GNRs (AGNRs), analyzed in terms of simple group theory and topological concepts and connections. All geometrical structures have been optimized (or reoptimized) using tight convergence criteria at the DFT level of the hybrid PBE0[26] functional using the 6-31G(d) basis set, as is implemented in the GAUSSIAN program package[27].The same package was also used for the calculation of NICS(1) aromaticity index, which for the present work has been proven satisfactory and suitable[2, 14-15]. This level of theory, used consistently and uniformly for all structures small and large (for all related properties), is fully adequate for such calculations, as was pointed out earlier.[1-2, 12-13]

# Supplementary Information

# Bridging the Physics and Chemistry of Graphene: From Hückel's Aromaticity to Dirac's Cones and Topological Insulators

Aristides D. Zdetsis*

Molecular Engineering Laboratory, Department of Physics, University of Patras, Patras

26500 GR, Greece

**<u>Disclaimer</u>**      This is not a full or complete study of *topological insulators* or of *aromaticity*; but the terms are used and discussed in the simplest possible context, using the most insightful, physical, pedagogical and transparent approach (according to the present author).

**#S1) Symmetry of π Molecular Orbitals of Benzene**

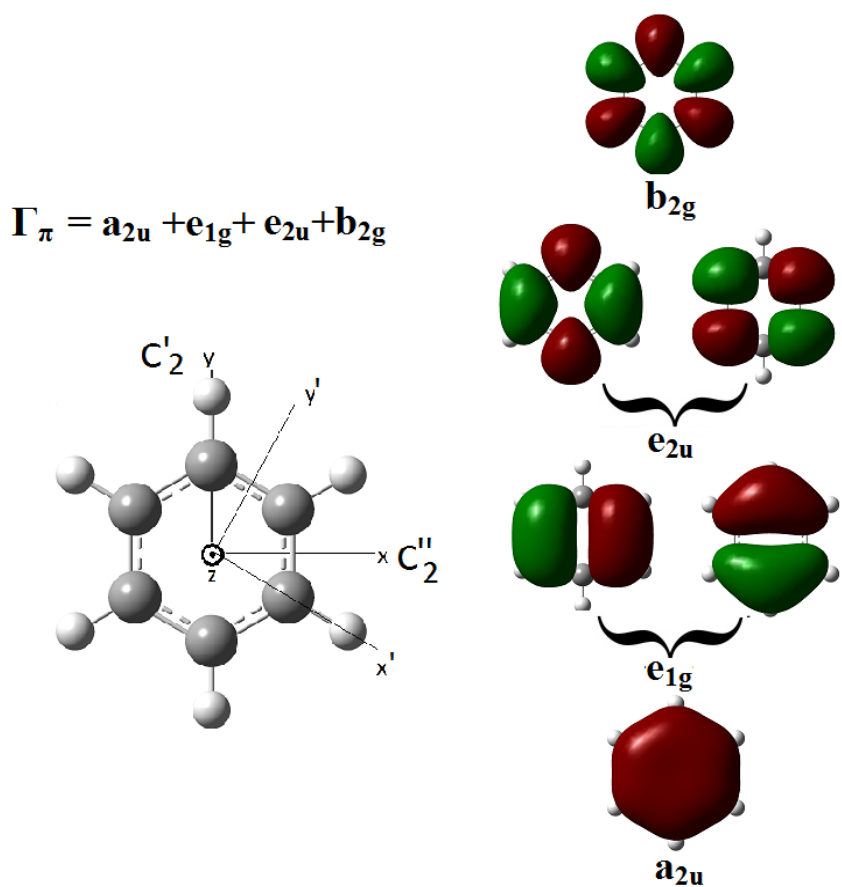

$$\Gamma_\pi = a_{2u} + e_{1g} + e_{2u} + b_{2g}$$

**FIGURE S1**. Some symmetry elements (left) and π-MOs of Benzene in order of increasing energy (from bottom to top) in analogy to Fig. 2 of the paper (right). The $C'_2$ rotation axis is directed trough the atoms, whereas the $C''_2$ axis directed through the bonds.

Observe that MOs of even parity (g) have opposite phases with respect to inversion at the origin (which interchanges sublattices). The opposite is true for odd parity (u) MOs. Also note that the doubly degenerate $e_{1g}$ and $e_{2u}$ representations describe "equivalent" pairs of MOs which are either symmetric or antisymmetric with respect to rotations about the $C''_2$ axes (and similarly for $C'_2$ axes).

**#2) Review of Shell Structure and Sublattice Symmetry for Hexagonal Samples.**

The first seven PAHs of the main sequence shown in Fig. 1 can be seen as the first seven individual "topological" shells of one PAH with "principal" shell number $n=7$ , as illustrated in Fig. S2(a). It has been shown[1] that (in analogy to the atomic shell model underlying the periodical table of the elements, described in terms of the atomic quantum numbers n, l, $m_l$ , s , and $m_s$)  the complete electronic configuration of the topological shell model can be as well specified by the corresponding "topological shell numbers"[1] $n$, $l$, $m_l$ , $s$ , and $m_s$ , where $n,$ the "principal shell number", denotes the total number of shells or layers. For a given PAH with a specified total shell number $n$, the shell model fully determines the complete configuration of the $\pi$ electrons, the aromatic and non-aromatic ("full" and "empty") rings defining the aromaticity pattern, as well as the symmetry and morphology of the frontier orbitals, i.e. the highest occupied and the lowest unoccupied molecular orbitals (HOMO and LUMO, respectively).

Also, it  has been shown[1] that the HOMO of the PAH with principal shell number $n$, PAH[$n$], can be written as a superposition of the LUMO of the "core" PAH[$n$-1] and the "surface" (or edge) HOMO orbital of the n[th] valence shell (annulene) ring. Similarly, the LUMO of PAH[$n$], can be written as a superposition of the HOMO of PAH[$n$-1] and the LUMO orbital of the n[th] valence shell (annulene) ring (see for instance Fig.3 in ref.1).

In other words, we can write[1]:

HOMO[PAH(n)]=LUMO[PAH(n-1)]+HOMO[Ann(n)]    (S1)

LUMO[PAH(n)]=HOMO[PAH(n-1)]+LUMO[Ann(n)]     (S2),

where Ann(n) indicates the annulene ring $C_{N_{A_{nn}}} H_{N_{A_{nn}}}$, with $N_{Ann} = 12n - 6$.

Obviously, this coupling goes on to the next PAH, i.e.:

HOMO[PAH(n+1)]=LUMO[PAH(n)]+HOMO[Ann(n+1)]    (S1')

LUMO[PAH(n+1)]=HOMO[PAH(n)]+LUMO[Ann(n+1)]     (S2')

**#3) Geometrical Shell and Sublattice Structures**

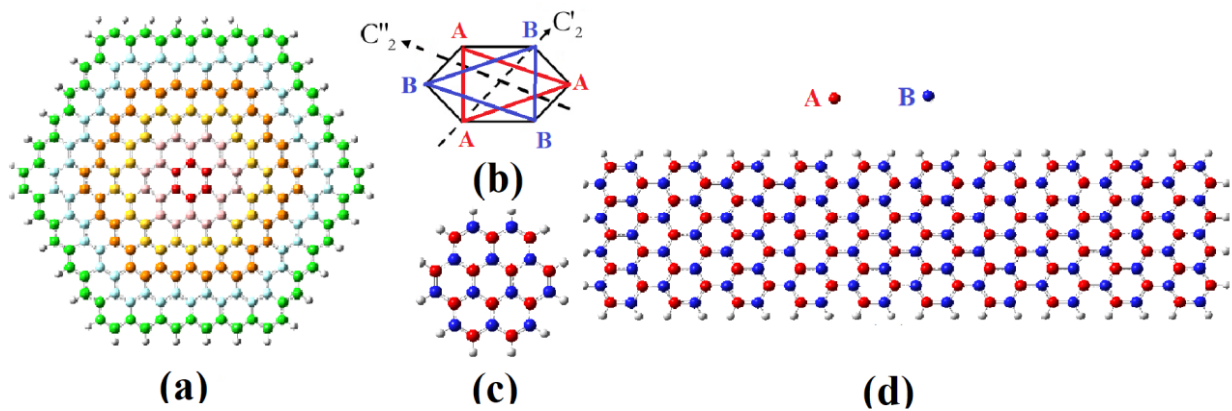

**FIGURE S2**. **(a)** The shell structure: Carbon atoms in different shells are colored with different color. The π electrons in each shell are fully characterized by "geometrical" quantum numbers $n$, $l$, $m_l$, $s$, and $m_s$ in full analogy to the corresponding atomic quantum numbers (see ref. 1). **(b)** The sublattice structure: A and B denote the sites of the two different sublattices. $C_2^{'}$ and $C_2^{''}$ are the two different types of $C_2$ rotation axis of Fig. S1 (through the atoms, and through the bonds, respectively) of the hexagonal $D_{6h}$ symmetry group. The sublattice structure of nanographene models with $D_{6h}$ hexagonal (**c**), and a of rectangular $D_{2h}$ symmetry (**d**). Carbon atoms belonging to different sublattices are colored differently. As is shown in the figure, the sublattice symmetry does not coincide with the full symmetry (space) group. For hexagonal nanographene(s) the sublattice symmetry is trigonal ($D_{3h}$); while for the rectangular ones is $C_{2v}$ (not $D_{2h}$).

It should be emphasized here that no topological edge states have been observed whatsoever in the HOMO and LUMO orbitals of the hexagonal samples we have examined in the present study. Nevertheless, loosely speaking, one can recognize some kind of topological features (e.g. geometrical phase or "Berry phase", but in real space) by considering two overlapping (infinitesimally touching) circles circumscribing the overlapping outer peripheries ($n$-1 and $n$ shells in the limit $n \rightarrow \infty$) with opposite parities of the wave-function. If one starts rotating on the periphery of the "interior" circle by $360^{\circ}$ and returns to the "same point" (infinitesimally close) on the "exterior" at equal energy, there will be a non-trivial phase difference of $\pi$ rad (or $180^{\circ}$), in the wave-function, which is in full accord with the ordinary meaning of geometrical (Berry's) phase (in real space).

### #3) Variation with size of the HOMO-LUMO energies and other properties

Figure S3 shows the variation of HOMO, LUMO energies, and gaps, and attempts to fit them for extrapolation beyond n=7. In the same figure it is shown that the average number of $\pi$ electrons per ring converges fast enough to the "crystalline" graphene value of 2, as it should be since each carbon atom contributes one $\pi$-electron, but belongs to three adjacent rings (6/3=2). The average number of $\pi$ electrons (per ring) can be easily calculated by recalling[1-2] that the total number of $\pi$ electrons in the PAH with shell number n is $N_C = 6n^2$, while the total number of rings is 1+6(1+2+…+n-1)=1+3n(n-1), since each layer $l$= 1, 2, 3, …, n-1 around the benzene nucleus contains $l$ benzene rings in each-one of its 6 sides. Thus, the average number of $\pi$ electrons (per ring) would be $6n^2/1+3n(n-1)$, which converges fast enough to the limit 2.

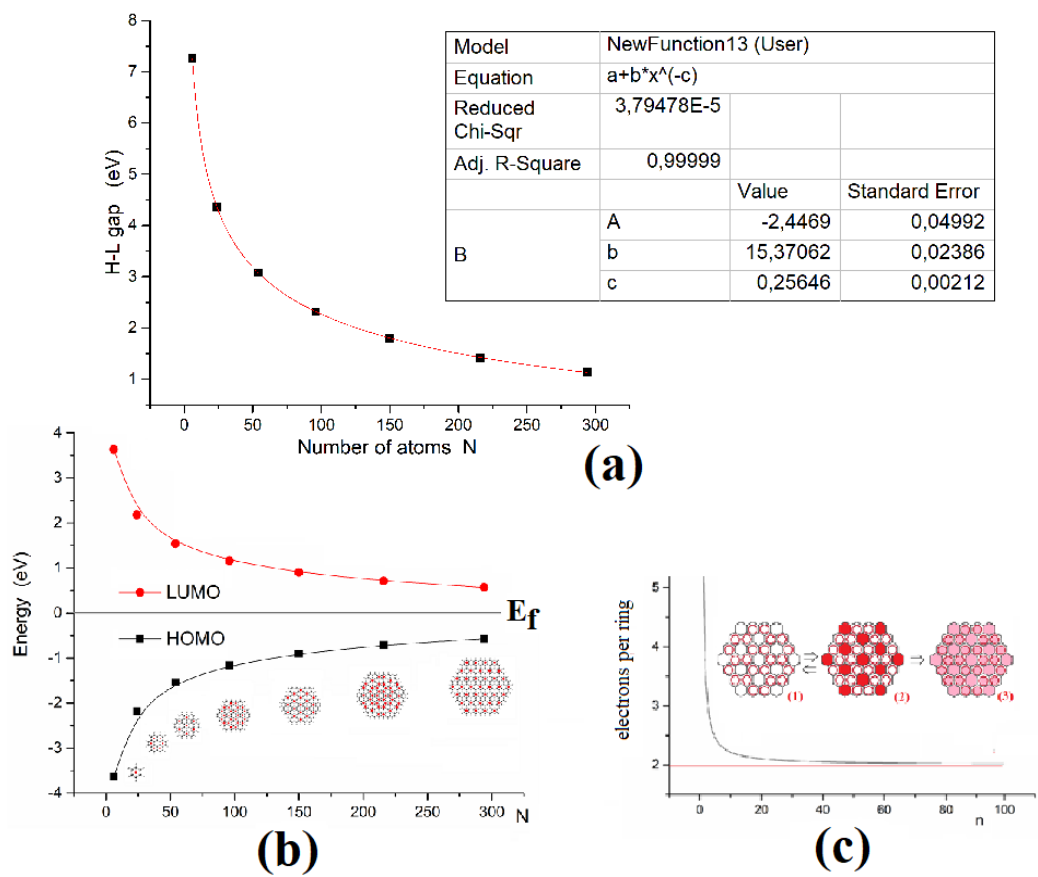

**FIGURE S3**. **(a)** HOMO-LUMO gap (top) as a function of the number N of carbon atoms (or π electrons). In the box it is also shown the results of a polynomial fit of the form: $E_g(N) = A + B \cdot N^{-\frac{1}{D}}$ , where A, B, and D = 1/C constants obtained from the fit, with D usually equal to the dimensionality of the samples[28] (here D≈4). **b)** HOMO and LUMO energies arranged symmetrically to the fermi energy level $E_f$ (in eV) for the PAHs of Fig. 1, as a function of the number of carbon atoms N. **(c)** Average number of π-electrons per ring as a function of the shell number *n* approaching (as n →∞) the value 2, as expected[9]. In the same figure the interchange of full and empty rings in graphene (1), (2); and the (time) average (3) of the two (see text) is shown schematically.

In the limit n →∞ (graphene) the "full" and "empty" rings should be dynamically interchanged since in this limit the results for n and n-1 or n+1 must coincide. This is shown schematically in Fig. S3(b) above: (1), (2); and the (time) average (3) of the two.

**#5) Synopsis of coupled or "interchanged" properties:**

<div align="center">

**TABLE S1**

</div>

Properties of "adjacent" PAHs of the main sequence (shell structure) and their extrapolation to graphene. Interchange is indicated by ⇌ and coupling by ∪ .

| | $n \rightleftharpoons$ n±1     **PAHs** | $n \rightarrow \infty$     **Graphene** |
|---|---|---|
| **1**<br><br>**Symmetry** | Trigonal ⇌Hexagonal | Trigonal $_A$ ∪ Trigonal $_B$ |
| **2**<br><br>**Sublattice** | Sublat. A (1,3,5) ⇌ Sublat. B (2,4,6) | Sublattice A ∪ Sublattice B |
| **3**<br><br>**Aromaticity** | CIRCO (Clar)⇌ CO (anti-Clar) | Clar ∪ anti-Clar |
| **4**<br><br>**Parity H, L** | $e_{1g}$ ⇌ $e_{2u}$ or g⇌u | g∪u |
| **5**<br><br>**Arom. Ring** | full rings ⇌ empty rings | full rings ∪empty rings |
| **6**<br><br>**Arom. Ring** | Aromatic rings along $C_2$'⇌ $C_2$''axis | Aromatic rings along $C_2$' ∪ $C_2$''axis |
| **7**<br><br>**Frontier**<br><br>**MOs** | HOMO ⇌LUMO | HOMO∪ LUMO |

| 8 | | |
|---|---|---|
| **Quasiparticle** | electron ⇌ hole | electron ∪ hole pairs |

## #6) Hexagonal PAHs with Armchair edges

As was described in ref. 1, for hexagonal PAHs beyond the "main sequence", of the form $C_\mu H_\nu$, such as the armchair PAHs of Fig. S4 below, one can define an effective shell number $n_{eff}$ through the relation[1] $n_{eff} = \nu/6$. The effective shell number is an integer since $\nu$, due to $D_{6h}$ symmetry, is always a multiple of 6.

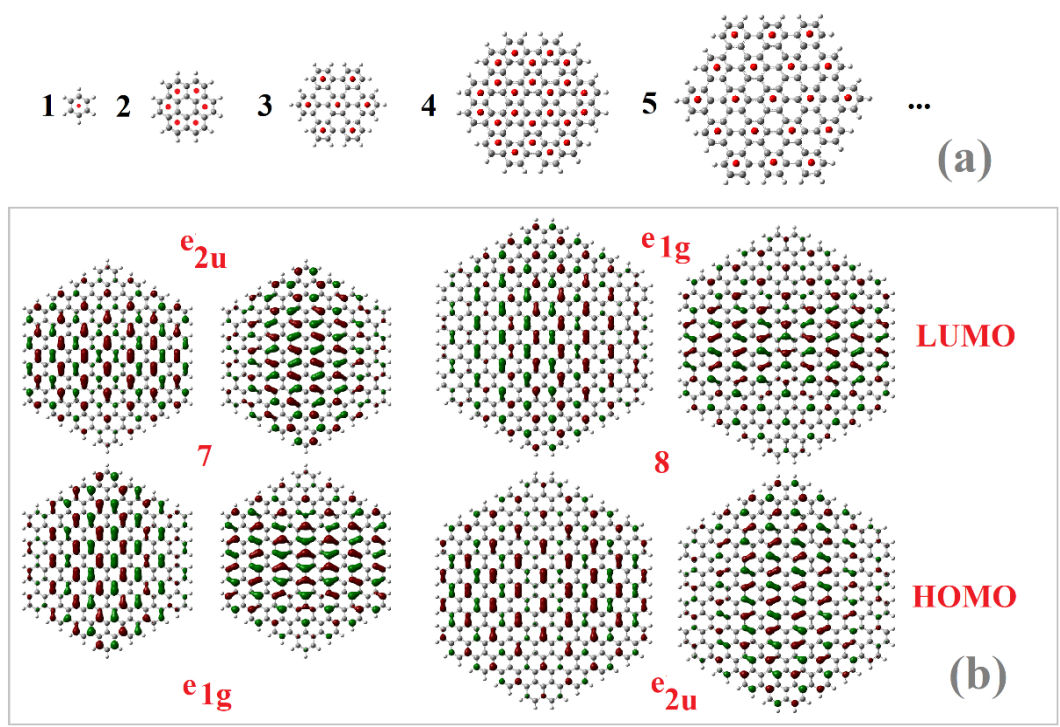

**FIGURE S4.** Armchair PAHs: (a) Aromaticity patterns for $n_{eff}$ =1, 2, 3, 4, 5,... ; (b) HOMO and LUMO MOs for PAHs with $n_{eff}$ = 7, 8.

The PAHs with $n_{eff}$ =1, 2, and 3 are benzene, coronene, and hexabenzocoronene, respectively. Benzene and coronene (with $n = n_{eff}$ =1, 2 respectively) are also members of the "main sequence".[1] Note that the fundamental shell model rules, relating $n$ or $n_{eff}$ with the type of aromaticity pattern, and the symmetry of HOMO, LUMO MOs,[1] are clearly valid for PAHs with armchair edges as well.

## #7) Rectangular Nanographenes and Nanoribbons

### #7.1) Aromaticity patterns

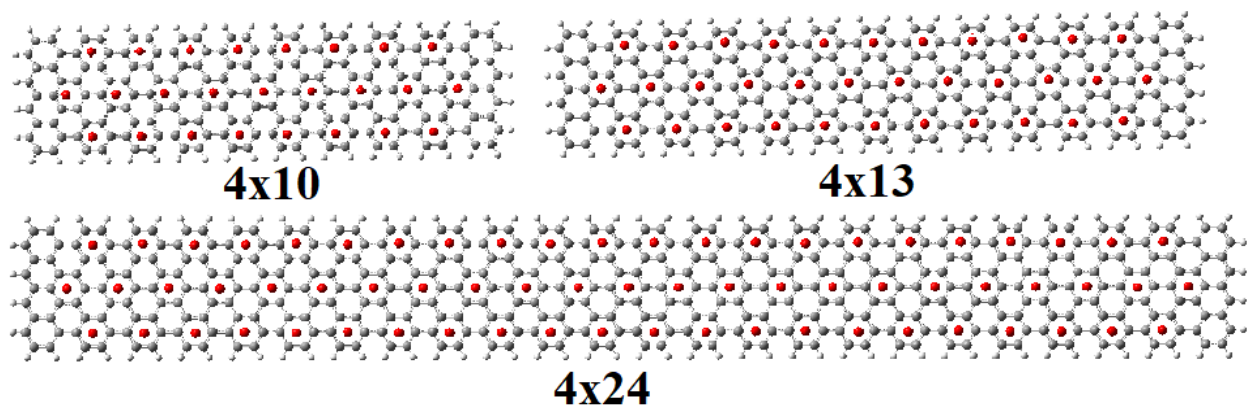

**FIGURE S5**. Aromaticity patterns of rectangular nanographenes, AGNRs, of constant width Z=4 (or in common notation W=9) for various lengths (number of armchair rings) A=10 (4x10), A=13 (4x13), and A=24 (4x24).

The aromaticity pattern (in this case, the characteristic Clar type CIRCO pattern) is independent of the length of the AGNRs, which are commonly characterized only by their width

### #7.2) Character table of the π electrons (in D$_{2h}$ geometry)

Note that the choice of the axis is different from Fig. S1. Now the $C_6$ rotation axis, which is perpendicular to the plane of the paper, is chosen as the x-axis, as is customary in group theory.

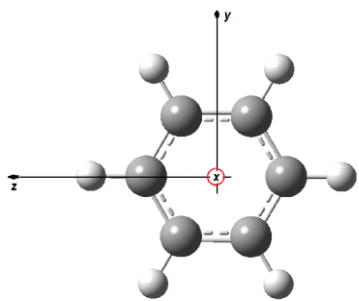

| $D_{2h}$ | E | $C_2(z)$ | $C_2(y)$ | $C_2(x)$ | i | $\sigma_{xy}$ | $\sigma_{xz}$ | $\sigma_{yz}$ |
|---|---|---|---|---|---|---|---|---|
| $\Gamma(6p\pi)$ | 6 | -2 | 0 | 0 | 0 | 0 | 2 | -6 |
| $B_{1g}$ | 1 | 1 | -1 | -1 | 1 | 1 | -1 | -1 |
| 2 $B_{2g}$ | 2 | -2 | 2 | -2 | 2 | -2 | 2 | -2 |
| $A_u$ | 1 | 1 | 1 | 1 | -1 | -1 | -1 | -1 |
| 2 $B_{3u}$ | 2 | -2 | -2 | 2 | -2 | 2 | 2 | -2 |

$D_{2h}$
$\Gamma(6p\pi) = B_{1g} + 2\,B_{2g} + A_u + 2\,B_{3u}$

**Scheme S1.** Group theory analysis of the $\pi$ electrons in $D_{2h}$ symmetry

Compatibility relations for the $D_{2h}$ symmetry group[17] show that the 2D hexagonal $e_{1g}$ and $e_{2u}$ $\pi$- MOs would be reduced as:

$e_{1g} \rightarrow b_{1g} + b_{2g}$ , and

$e_{2u} \rightarrow a_u + b_{3u}$

Furthermore, the 1D $b_{1g}$ and $a_{2u}$ $D_{6h}$ representations would be reduced in $D_{2h}$ as:

$b_{1g} \rightarrow b_{2g}$ ,

and

$a_{2u} \rightarrow b_{3u}$

### #7.3) Density of States (DOS) for $D_{2h}$ geometry NGRs and GNRs)

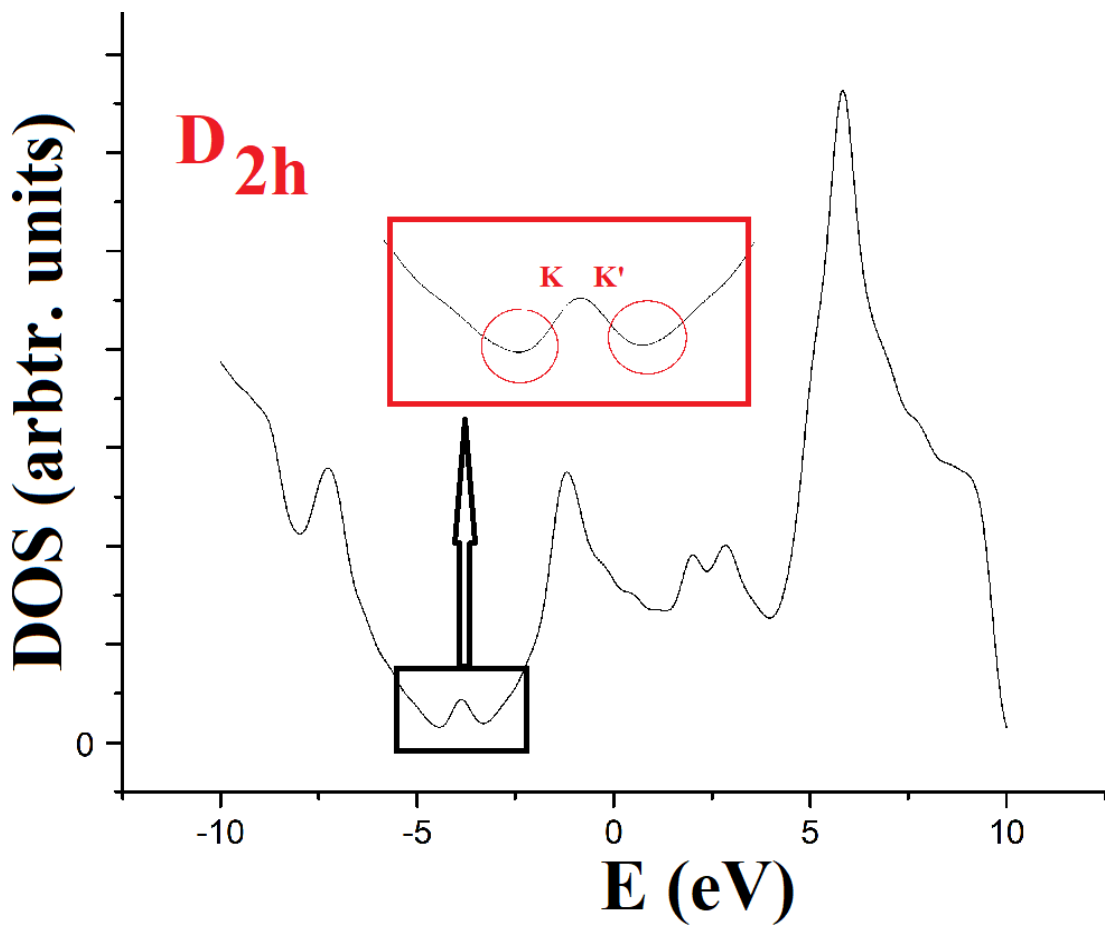

**FIGURE S6**. Simulated DOS (obtained by a gaussian broadening by 0.30 eV) of the energy levels) of the 15x15 rectangular NGR. The inset shows the region around the Fermi level and the corresponding K points, as edge states inside the gap. Compare with Fig. 2.